\documentclass[11pt]{article}
\usepackage[margin=1in]{geometry}
\usepackage{times}
\usepackage{latexsym}
\usepackage{amsmath}
\usepackage{amsthm}
\usepackage{float}
\usepackage{graphicx}
\usepackage{url}
\usepackage{afterpage}
\usepackage{footmisc}

\begin{document}
\title{A Survey of Fault-Tolerance and Fault-Recovery Techniques in Parallel Systems}
\author{Michael Treaster\\
National Center for Supercomputing Applications (NCSA)\\
University of Illinois\\
Email: treaster@ncsa.uiuc.edu}
\date{}
\maketitle

\begin{abstract}

Supercomputing systems today often come in the form of large numbers of commodity systems linked together into a computing \emph{cluster}.  These systems, like any distributed system, can have large numbers of independent hardware components cooperating or collaborating on a computation.  Unfortunately, any of this vast number of components can fail at any time, resulting in potentially erroneous output.  In order to improve the robustness of supercomputing applications in the presence of failures, many techniques have been developed to provide resilience to these kinds of system faults.  This survey provides an overview of these various fault-tolerance techniques.

\end{abstract}

\section{Introduction}

Computational tasks continue to become more complex and require increasing amounts of processing time.  At the same time, high performance computer systems are composed of increasing numbers of failure-prone components.  The end result is that long-running, distributed applications are interrupted by hardware failures with increasing frequency.  Additionally, when an application does fail, the cost is higher since because more computation is lost.  It is imperative that both distributed applications and parallel systems support mechanisms for fault-tolerance to ensure that large-scale environments are usable. 

In this paper, we focus specifically on cluster systems and the applications that run in these environments.  Cluster systems are typically composed of a large number of identical, centrally managed computation nodes constructed from commodity components and linked by one or more network infrastructures such as Ethernet or Myrinet.  Nodes employ software such as MPI to facilitate their integration into a larger, unified, cluster system.  Often, some nodes are dedicated to management, user interaction, and/or storage.  

In this environment, there are many opportunities for failure.  Any component in any compute node could fail.  This includes, but is not limited to, the processor, disk, memory, or network interface on the node.  A hardware or software failure on a management node could affect the entire system as scheduling and synchronization data is lost.  Failures external to individual nodes are also possible.  Many possible failures could remove a large number of nodes from the system, such as an air conditioning failure or a network switch failure.  Any of these failures will cause the applications running on the affected nodes to crash or produce incorrect results.  An overview of the theoretic models used to describe these types of faults and others is provided in Section~\ref{faultmodels}.

There are two fundamental classes of faults that can occur in cluster systems.  First, a centralized component such as a storage node or management software can fail as a result of a software bug or a hardware fault.  These centralized components are typically few in number, spanning only a tiny percentage of cluster nodes.  Protecting against these failures typically involves redundancy.  Critical functionality is replicated over several nodes such that if one node fails, a backup can step in to take over the responsibilities of the primary.  We discuss techniques for handling failures of centralized components in Section~\ref{centralized}. 

The second class of failure is a crash or hang of software on one of the many computation nodes in the cluster.  This can result from a software bug in an application, a hardware fault on the node, or a problem in the operating system local to the node.  All of these failures have the same end result:  the application running on the node can no longer function properly, but the other nodes participating in the same computation can continue unaffected excepting that they will no longer receive output from the failed node.  The standard technique for handling application failures is to periodically checkpoint the computational state of the application such that it can be restored in the event of a system failure.  Other nodes participating in the computation may need to be rolled back to earlier checkpoints in order to make them consistent with the recovery state of the failed node.  More detail on rollback recovery techniques and other methods of protecting against failures on compute nodes are provided in Section~\ref{applications}.

\section{Fault Models}
\label{faultmodels}

There are countless ways in which computing systems and applications may fail.  These failures can be categorized by abstract models that describe how a system will behave in the presence of faults.  A fault tolerance technique will assume a certain model of failure when making claims about the types of faults it can handle.  We present the two most common failure models here, as well as a relatively new model which is a more accurate, but more complex representation of how real systems work.   

\subsection{Byzantine Faults}

The Byzantine fault model represents the most adversarial model of failure.  This fault model allows failed nodes to continue interacting with the rest of the system.  Behavior can be arbitrary and inconsistent, and failed nodes are allowed to collude in order to devise more malicious output.  Correctly operating nodes cannot automatically detect that any nodes have failed, nor do they know which nodes in particular have failed if the existence of a failure is known.  This model can represent random system failures as well as malicious attacks by a hacker.  It has been proven that no guarantees can be made concerning correct operation of a system of $3m+1$ nodes if more than $m$ nodes are experiencing Byzantine failures\cite{lamport82byzantine}.

\subsection{Fail-stop Faults}

The fail-stop fault model is much simpler than the Byzantine model.  This model allows any node to fail at any time, but when the failure occurs it ceases producing output and interacting with the rest of the system.  Furthermore, all other nodes automatically know that the node has failed.  This fault model represents common modes of failure such as a system hang or crash, but does not handle more subtle failures such as random memory corruption\cite{schlichting83}.   

\subsection{Fail-stutter Faults}

The Byzantine fault model is extremely broad, perhaps unrealistically so, and is therefore extremely difficult to analyze or tolerate.  The fail-stop model is commonly used when presenting fault-tolerance techniques, but it is often criticized as overly simplistic due to its failure to represent many types of real-world failures.  The fail-stutter fault model\cite{arpaci01fail} is an attempt to provide a middle ground model between these two extremes. 

The fail-stutter model is an extension of the fail-stop model.  It attempts to maintain the tractability of that model while expanding the set of real-world faults that it includes.   The fail-stutter model includes all provisions of the fail-stop model, but it also allows for \emph{performance faults}.  A performance fault is an event in which a component provides unexpectedly low performance, but to continues to function correctly with regard to its output.  This extension allows the model to include faults such as poor latency performance of a network switch when suddenly hit with a very high traffic load.  Despite its advantages, however, this fault model has yet to see widespread acceptance by the community.

\section{Fault-Tolerance of Centralized Components}
\label{centralized}

Cluster systems depend on many centralized components to function properly.  Management nodes handle job scheduling and node monitoring responsibilities.  Storage nodes provide access to high-capacity disk arrays.  Head nodes allow users to interact with the system without infringing on processor time on the compute nodes.  Each of these components typically appears in small numbers in a cluster, and they are dwarfed by the number of compute nodes.  Due to the critical importance of these components and the small fraction of the total system they compose, it is feasible to dedicate extra attention to these resources and possibly allocate additional hardware to ensure they are robust in the event of system failures.

\subsection{Replication}

The most common means of providing fault-tolerance for centralized system components is to replicate the functionality.  This replication can take one of two forms.  

In \emph{active replication}, a second machine receives a copy of all inputs to the primary node and independently generates an identical system state by running its own copy of all necessary software.  Additionally, the backup node monitors the primary node for incorrect behavior.  In the event that unexpected behavior is observed (such as a system crash), the backup node promotes itself to primary status and takes over the critical functionality for the system.  Since its system state is already identical to that of the primary, this changeover requires a negligible amount of time.  This type of replication is infeasible for compute nodes because doubling the number of these nodes would increase the cost of the system by nearly double without increasing the computational capacity of the system\cite{shokri97roafts, goldberg01design}.

A variant on active replication calls for multiple backup systems for the primary.  All replicas receive all input messages.  When output is generated, all replicas compare their results using a Byzantine algorithm in order to vote on what the correct output should be.  If one or more nodes generate incorrect output beyond a threshold number of times, they are marked as faulty and ignored until they can be repaired by maintenance procedures.  This variant is capable of handling Byzantine faults, while the previous implementation of active replication can only tolerate faults in the fail-stop model\cite{reiter95rampart, li01faulttolerant}.  

In \emph{passive replication}, a ``cold spare'' machine is maintained as a backup system to the primary.  This system typically resides in an idle or powered off state, but it has a copy of all necessary system software used by the primary.  If the primary machine should fail, the cold spare takes over control of the primary's responsibilities.  This may incur some interruption of service, depending on the idle state of the replica.  Unless additional checkpointing functionality is employed, this type of replication is not suitable for components that maintain complex internal state that cannot be easily recovered.  However, it is appropriate for components that maintain minimal internal state, such as monitoring subsystems\cite{leangsuksun03building}.   

It is also possible to replicate components at a finer level than by simply duplicating an entire machine.  It is possible to construct machines with redundant internal components.  For example, a machine could be equipped with two processors, two memory banks, two motherboards, and so forth, possibly targeting only the most failure-prone components.  By carefully monitoring the correctness of each component, a failing component can be automatically disabled and replaced with the backup without bringing down the system and without interrupting any system software.  Once correct functionality is restored, the failed component can be hot-swapped with a new copy by an administrator at a later time\cite{hough00algorithm, stratus}.

\subsection{Reliable Communication}

Active replication of system components is viable only if all replicas are guaranteed to receive exactly the same inputs.  Although communication with centralized nodes can be implemented with some kind of multicast protocol, there is not necessarily any guarantee that the multicast is perfectly consistent across all recipients.  If this guarantee is not provided, replicas can end up out of sync if a transient network error results in a message being transmitted differently to different replicas. 

There are a variety of multicast protocols that exist which provide various guarantees on reliability, security, and ordering.  Since these protocols are not the focus of this survey, we presented only two as examples of how these protocols might work.  

One multicast protocol\cite{moser96totem} uses a token-based system.  A virtual token is passed from node to node, and only the holder of the token is allowed to send messages.  This prevents multiple simultaneous messages from being sent and interfering with message ordering.  Messages are forwarded from one node to the next around a ring network overlay, ensuring that all nodes receive all messages if no error occurs.  To address fault-prone links, Each message is marked with a sequence number.  If a node does not receive a particular message but does receive a later message, it can detect the missing message by looking for gaps in the sequence numbers off received messages.  In this event, the node can request retransmission of the earlier message.  

This protocol allows multiple such rings, each with its own token, to be joined together.  In this case, messages are also tagged with a timestamp generated by a Lamport clock\cite{lamport78}.  Gatekeeper nodes that participate in multiple rings use the timestamps to translate the sequence number from the originating  ring to the appropriate sequence number for the destination ring.  Message transmission can then work as in the single-ring case, using sequence numbers to order messages even when multiple tokens are used.  

Another approach\cite{reiter95rampart} separates multicast atomicity from multicast reliability.  In this protocol, atomicity is implemented by using a \emph{sequencer} node.  The sequencer can specify ordering in one of two ways.  In the first technique, all messages are sent to the sequencer, which then multicasts the message to all intended recipients.  Since the sequencer acts as an intermediary in this scenario, it is able to impose an order on all messages.  In the second technique, a sender node multicasts a message directly to all recipients, and also sends the message to the sequencer.  The sequencer then multicasts a second message to all recipients of the first message indicating the order in which the first message should be received.  Reliability in this protocol is ensured using a Byzantine agreement protocol to guarantee that all recipients receive the same message.

\subsection{Monitoring}

For a system to automatically activate a backup replica, it must detect that a primary component has failed.  There are a variety of monitoring approaches that are commonly used.  

The most basic form of monitoring is a simple heartbeat system.  A monitor process listens for periodic messages from the monitored components.  The message simply indicates that the component continues to function correctly enough to send messages.  If the monitor fails to receive a message from a component within a certain threshold of time, it notes the component as failed.  This form of monitoring is suitable for detecting failures in a fail-stop model.  More subtle faults, such as erroneous computation, are not detected by heartbeats\cite{leangsuksun03building, kalbarczyk99chameleon}.  Variations on heartbeats also exist in which messages are not all sent to a single monitor, but are instead randomly gossiped throughout the system or are forwarded up a hierarchy of nodes.  These methods allow heartbeating to scale across larger numbers of nodes, and they are typically used in systems with a more distributed nature\cite{hough00algorithm}.

Byzantine consensus among component replicas is another form of monitoring.  The replicas vote on the correct output or action based on the observed inputs.  In the event of a disagreement, the minority is considered faulty.  This method allows for the detection of Byzantine faults, resulting in a much more powerful monitor.  However, the communication costs of Byzantine consensus are much higher than the costs for heartbeat monitoring, since in the worst case all nodes must communicate with all other nodes.  However, for a small number of replicas, this is likely to be acceptable\cite{reiter95rampart, lamport82byzantine, goldberg01design, bottomley02clusters}.  

Monitoring can also be implemented in the form of a self-consistency check.  A node can periodically run diagnostics in order to determine whether or not internal components are operating correctly.  Additionally, other nodes in the system can share information in order to determine if a centralized server might be experiencing a problem.  Other nodes can also detect errors by checking for missing or malformed messages from the server.  In the event that other nodes suspect a problem exists, they can trigger a self-diagnosis check on the server\cite{stratus, goldberg01design}.

\subsection{Software Engineering}

Management software can be designed with fault-tolerance in mind.  Careful software engineering can produce programs that are resistant to system faults, and that minimize down time when fault does impair functionality of the system.  

Restarting a system is common way of dealing with a wide variety of faults.  Unfortunately, this can incur significant down time, and as a side effect it can also result in the restarting of correctly functioning components.  One way the impact of faults is by dividing software into many small, independent components.  These components are organized in a hierarchy, with modules higher in the hierarchy drawing functionality from the lower levels.  If a fault occurs in one of these components, those components that are directly affected can be restarted while the rest of the system continues functioning.  This allows a system to handle faults at multiple levels while minimizing impact on other aspects of the system.  For example, a web server could be dependent on a database, which is then dependent on a machine.  If the web server experiences an error, it should be restarted individually without interfering with the database.  By extending this idea as far as possible, only small pieces of the system will need to be restarted to cope with any given error\cite{candea01recursive, candea04improving}.   

\section{Fault-Tolerance of Parallel Applications}
\label{applications}

The real value of a cluster system is not in the management software that allows it to operate, but rather in the execution of large-scale parallel applications that would require intractable amounts of time to run in a serial environment.  Therefore, even if the management nodes on a cluster are completely resistant to any kind of fault, the compute nodes and the applications that run on them must also be protected or the efforts on the management nodes are meaningless.

\subsection[Checkpointing and Rollback Recovery]{Checkpointing and Rollback Recovery\footnote{Much of the information contained in this section is drawn from a larger survey of checkpointing and rollback recovery techniques by Elnozahy \emph{et al}\cite{elnozahy-survey}.  The reader is directed to this work for more complete information on the topic of rollback recovery techniques.}}

The most basic form of fault tolerance for parallel applications consists of checkpointing and rollback recovery.  To ensure application reliability, the need to preserve an application's state in order to preserve completed computation in the event of a system failure is extremely important.  Rollback recovery techniques are a common form of this type of state preservation, and they have received a great deal of attention from the research community.  

Rollback recovery techniques model a message-passing application as a fixed number of processes in a distributed system that communicate over a network by sending messages.  Processes are assumed to have access to some kind of \emph{stable storage} that will survive even in the event that the process fails.  Periodically during the execution of an application, the system records to stable storage a snapshot of processes composing the application.  In the event that a process fails, the application's computational state can be restored to some fault-free state by rolling back all processes to the most recent checkpoint state\cite{elnozahy-survey}.

A process is modeled as a sequence of \emph{state intervals}.  Each state interval begins with a nondeterministic event, such as a user input or the arrival of a message, but then consists of deterministic execution until the next state interval is reached. \cite{elnozahy-survey}

Checkpointing and rollback recovery techniques generally assume a fail-stop model of faults.  It is assumed that each process has access to some kind of stable storage, which can still be accessed after the process has failed.  ``Stable storage'' can come in a variety of forms, depending on the assumptions made by the recovery protocol.  It need not be an actual disk.  If a system must tolerate only a single failure, stable storage could be implemented by using the volatile memory of other processes in the system.  If failures are assumed to be transient, the local hard disk of a process host may be used\cite{borg89, johnson87}.  If failures are not transient, the local hard disk of a host cannot be used, since that disk will not be accessible after a host failure.  In this case, stable storage for a process must be found remotely with respect to the process host\cite{lamson79}.

A set of checkpoints is \emph{consistent} if, for each message that a checkpoint registers as being received another checkpoint records having sent the message.  In other words, there is no message which, according to the set of checkpoints, was received but never sent.  A set of checkpoints can be (but is not necessarily) inconsistent if a receiving process writes a checkpoint later (in real time) than a sending process\cite{chandy85}.  

To recover from a process, the recovery procedure must ensure that the internal state of the recovered process conforms to the observed state of the system before the failure.  This is accomplished by identifying the most recent set of consistent checkpoints and restoring the system to the state recorded in this set.  This set of checkpoints is called the \emph{recovery line}\cite{randell75}. 

Rollback recovery techniques can be subdivided into two broad categories:  checkpoint-based, and log-based.  

\subsubsection{The Domino Effect}

The \emph{domino effect} occurs when, after a rollback, one process has recorded the receipt of a message that has not yet been sent according to the state of the sender.  In this case, the recipient's state must be rolled back to a state prior to the receipt of the message.  This effect can cascade back through the computation history indefinitely, resulting in the loss of large amounts of computation.  This effect also forces each process to maintain multiple checkpoints, such that fallback options exist in the event of a domino effect rollback\cite{randell75}.

\subsubsection{Checkpoint-based Rollback Recovery}

In general, checkpoint-based rollback recovery protocols periodically save the current computational state of each process involved in a computation.  When checkpoint creation is signaled, the system records a representation of the memory state of the process, such that the process state can be reconstructed at an intermediate state in the computation if the process should fail.  

There are three subcategories in the category of checkpoint-based recovery protocols:  uncoordinated checkpointing, coordinated checkpointing, and communication-induced checkpointing.

In \emph{uncoordinated} (or asynchronous) checkpointing schemes, each process decides when to take its own checkpoints.  This relieves synchronization complexity and allows processes take checkpoints when it is most convenient or efficient to do so.  This approach does incur several disadvantages, however.  First, a process may decide to take a checkpoint that cannot or will not be used in a consistent global state, resulting in unnecessary overhead in the creation of the checkpoint.  Second, using the checkpoint may result in the domino effect.  Most research in this class of checkpointing focuses on determining the set of checkpoints to use when restoring an application state after a failure\cite{elnozahy-survey}.  

\emph{Coordinated} (or synchronous) checkpointing techniques force processes to organize their checkpoints such that together they generate a single, consistent application checkpoint.  This can be accomplished by having a checkpoint initiator process send a message to all processes at a particular time to force all processes to checkpoint roughly simultaneously.  This increases the complexity of generating the checkpoints, but it reduces the complexity of restoring the application state from a checkpoint because there is no need to search for a consistent slice of checkpoints.  Additionally, it avoids the domino effect, since processes plan their checkpoint times specifically to avoid problems such as this.  Finally, since the domino effect is avoided each process needs to maintain only a single checkpointed state, which can be replaced when a new checkpoint is recorded\cite{elnozahy-survey}.   

Coordinated checkpointing suffers from consistency problems because each process receives the initiation message at only approximately the same time instead of at exactly the same time.  Process $p_0$ might receive the checkpoint message first, complete the checkpointing, resume computation, and send a message to $p_1$ before $p_1$ receives the checkpointing message.  In this event, $p_1$ might record in its checkpoint the receipt of the message from $p_0$, but $p_0$ will not record having sent it.  To address this situation, $p_0$ can send a checkpoint initiation message ahead of the first message after it records a checkpoint (assuming FIFO message ordering), or it can piggyback the checkpoint request onto the first message\cite{elnozahy94}.

If all processes have loosely synchronized clocks, coordinated checkpointing can be accomplished without the use of special checkpoint signal messages.  Each process automatically checkpoints at a pre-specified time $t$, then waits to ensure no failures occurred.  After the timeout period expires, the process knows that no failures have occurred and it can continue processing\cite{cristian91, tong92}.

\emph{Communication-induced} (or quasi-asynchronous) checkpointing provides resilience to the domino effect without requiring global coordination across all checkpoints.  Each process takes checkpoints locally, as in uncoordinated checkpointing.  However, such protocols allow for processes to be forced to take a checkpoint in order generate a global checkpointed state that will not succumb to the domino effect.  Each message passed between processes contains extra protocol information that allows the recipient to determine for itself whether or not it should take a forced checkpoint\cite{elnozahy-survey}.

\subsubsection{Log-based Rollback Recovery}

Log-based rollback recovery protocols, or message logging protocols, supplement normal checkpointing with a record of messages sent by and received by each process.  If the process fails, the log can be used to replay the progress of the process after the most recent checkpoint in order to reconstruct its previous state.  This has the advantage that process recovery results in a more recent snapshot of the process state than checkpointing alone can provide.  Additionally, log-based approaches avoid the domino effect, since the failed process can be brought forward to the global application state rather than individual processes being forced to roll back for consistency with the failed process. 

Log-based recovery protocols rely on the \emph{piecewise deterministic assumption} (PWD).  This assumption dictates that the system has the ability to detect the nondeterministic events that transition to the next state interval.  Furthermore, the system must be able to record information about the events such that the important aspects of the event can be recreated in a reconstruction of the process state\cite{strom85}.   

An orphan process $p$ is a process that does not fail, but whose state depends on a nondeterministic event that was not recorded to stable storage and the determinant of which was not recorded on $p$.  Such a process therefore cannot be restored to a consistent state because the information required to replay an event has been lost\cite{elnozahy-survey}. 

There are three main techniques used by log-based recovery protocols to guarantee that all processes can be recovered to a consistent state in the event of a failure:  Pessimistic, optimistic, and causal.  Each of the three approaches has its own tradeoffs for performance, ease of process recovery, and ability to roll back processes that did not fail.

\emph{Pessimistic} logging techniques, sometimes called synchronous logging, record the determinant of each event to stable storage before the event is allowed to affect the computation.  This ensures that the system will easily be able to recover from the failure of any process occurring at any time, because no process can be affected by an event that has not been logged.  Pessimistic logging has two key advantages.  First, in the event of a failure, processes that did not fail can never become orphans and need not take any special actions.  This greatly simplifies the recovery algorithm.  Second, garbage collection of message logs and checkpoints is simple - only one checkpoint must be maintained for each process, and message logs older than that checkpoint can be discarded\cite{elnozahy-survey}.

\emph{Optimistic}, or asynchronous, logging techniques record logs to volatile storage, which is then periodically written to stable storage.  This substantially reduces the performance overhead on the application because it does not need to block while waiting for each message to be written to disk.  Unfortunately, recovery of the system in the event of a failure is much more complex.  Since messages recorded in volatile memory will be lost in the event of a process failure, processes can become orphans.  In addition to the recovery of the failed processes, the surviving processes must be rolled back to a state that does not depend on any lost messages\cite{strom85}.

\emph{Causal logging} protocols maintain the advantages of both optimistic and pessimistic logging, but at the expense have requiring much more complex recovery techniques.  The low overhead of optimistic logging is attained by saving logs to volatile storage, similar to optimistic logging\cite{alvisi96, elnozahy94}.

\subsection{Virtual Processors}

Virtual processors allow parallel application execution to be subdivided across a larger number of processors than physically exist.  In such systems, each virtual processor is mapped to a physical processor, and each physical processor handles one or more virtual processors.  Virtual processors can be migrated from one physical processor to another without interfering with the overall execution off the application.  This technology has applications to load balancing and grid computing, and also to fault tolerance.  

The key to the fault tolerance application is the migration feature.  If a computation node should suffer impaired performance, the virtual processors can be migrated to other nodes that remain fully operational.  Although the execution speed of the application as a whole would be slightly decreased due to the load being shared over fewer processors, this is a far better solution than the application failing completely, or pausing until a new node becomes available for a traditional checkpoint-based recovery solution to be effected.  

To handle the fail-stop class of failures, the virtual processor techniques can be combined with processor replication or checkpointing to allow for the run-time recovery of the lost virtual processors on a surviving physical node.  In practice, this is done using coordinated checkpointing techniques, using in-memory checkpoints for small applications or disk-based checkpointing for large applications\cite{ftpds04}. 

Alternatively, virtual processors can be actively or passively replicated to provide backup instances that can take over in the event of a failure.  Different aspects of application functionality can be assigned to each processor, and processors can be replicated to different degrees depending on the level of protection that is required for that component\cite{powell93distributed}.  

\section{Implementation/Packaging}

Although fault tolerance techniques can have academic merit in the research lab, it is only through deployment that they attain real value to application developers.  Obviously fault tolerance techniques can be implemented directly by the application developer, but Providing fault tolerance in a simple, easy-to-use package lowers the learning curve of the software and encourages adoption of new techniques.  Therefore, the interface to these systems can be a key differentiating factor between various fault tolerance solutions.  

The most common packaging for a fault tolerance technique is in the form of a software library that provides an application programming interface (API) to the library.  The developer calls specific API functions at certain points in the application source code to use functionality from the library.  A library that requires less modification to the application code is generally easier to use and will see more widespread adoption among developers\cite{birman85replication, badrinath02ftop, blochinger00dependable}.  Some APIs are object-oriented, and the developer instantiates certain objects that implement fault-tolerant properties\cite{arjuna}.  

A variation on a library API is the idea of meta-objects.  Like an object-oriented library, a meta-object library implements objects with fault tolerant properties.  However, instead of instantiating objects directly, the developer extends these classes when specifying classes for the application.  This causes the application objects to inherit fault tolerance functionality from these other objects\cite{fabre98metaobject}. 

It is also possible to build fault-tolerant programming languages.  \cite{avalon} describes an extension to the C++ programming language.  It simplifies the implementation of a distributed application by hiding its distributed nature from the developer, thereby reducing programming errors that could result in runtime errors.  Additionally, it provides high-level language support for synchronization and fault-tolerance properties.  

Fault tolerance can also be implemented at an operating system level.  In the best case, no changes are required for applications to take advantage of the capabilities.  They are automatically protected simply by virtue of running on the enhanced system.  In other cases, applications must be aware of the fault tolerant capabilities of the system and make certain system calls to make use of them.  From the perspective of the developer, this is similar to using a software library, but the protection afford can sometimes be greater due to the integration with the operating system\cite{powell93distributed}.

Another possible approach is to deploy system middleware, an infrastructure of system daemons, to monitor and support applications.  This is similar to integrating fault tolerance at the system level, but it does not require a special host operating system.  Instead, system-level daemon software runs continuously in the background.  These daemons can be configured to monitor any or all applications and provide them with any of a variety of fault tolerance techniques.  Additionally, the same techniques can be applied to the fault tolerance infrastructure itself to ensure that it continues functioning properly\cite{kalbarczyk99chameleon}.

The last method of delivering fault tolerance techniques is to use a preprocessor to automatically generate the necessary modifications to the application source code.  This can simplify the process of making complex code changes to an application.  In some approaches, the developer must mark certain code locations to indicate good places for fault tolerance code to execute, such as a code region with a small memory footprint for optimal checkpointing.  In other cases, the preprocessing is entirely automatic\cite{schulz04implementation, pingali04application}.

\section{Conclusions}

This survey presents an overview of fault-tolerance techniques in large scale cluster computing systems.  These techniques can be grouped into two categories:  protection for the cluster management hardware and software infrastructure, and protection for the computation nodes and the long-running applications that execute on them.  

Cluster management hardware and software fault-tolerance typically makes use of redundancy, due to the relatively small number of components that need to be duplicated for this approach.  When a component fails, the redundant components take over the responsibilities of the failed parts.  Redundancy can also be used for fault detection by comparing the outputs produced by each replica and looking for discrepancies.  

Cluster applications are protected from faults using checkpointing and rollback recovery techniques.  Each process cooperating in the application periodically records its state to a checkpoint file in reliable, stable storage.  In the event of a process failure, the application state is restored from the most recent set of checkpoints.  There are a variety of protocols that have been developed to determine when processes should record checkpoints and how to restore the application state.  

Fault tolerance solutions can be implemented in a variety of forms.  This include software libraries, special programming languages, compiler or preprocessor modifications,  operating system extensions, and system middleware.  Each method has its own tradeoffs in terms of power, portability, and ease of use.

\small
\bibliographystyle{abbrv}
\bibliography{treaster-clusters-corr}

\begin{thebibliography}{10}

\bibitem{alvisi96}
L.~Alvisi.
\newblock {\em Understanding the message logging paradigm for masking process
  crashes}.
\newblock PhD thesis, Cornell University, Department of Computer Science, 1996.

\bibitem{arpaci01fail}
R.~H. Arpaci-Dusseau and A.~C. Arpaci-Dusseau.
\newblock Fail-stutter fault tolerance.
\newblock In {\em Proceedings of the Eighth Workshop on Hot Topics in Operating
  Systems}, page~33. IEEE Computer Society, 2001.

\bibitem{badrinath02ftop}
R.~Badrinath, R.~Gupta, and N.~Shrivastava.
\newblock {FTOP}: A library for fault tolerance in a cluster.
\newblock In {\em 14th IASTED Intl. Conf. on Parallel and Distributed Computing
  and Systems}, November 2002.

\bibitem{birman85replication}
K.~P. Birman.
\newblock Replication and fault-tolerance in the isis system.
\newblock In {\em Proceedings of the tenth ACM symposium on Operating systems
  principles}, pages 79--86. ACM Press, 1985.

\bibitem{blochinger00dependable}
W.~Blochinger, R.~B{\"u}ndgen, and A.~Heinemann.
\newblock Dependable high performance computing on a parallel sysplex cluster.
\newblock In {\em PDPTA}, 2000.

\bibitem{borg89}
A.~Borg, W.~Blau, W.~Graetsch, F.~Hermann, and W.~Oberle.
\newblock Fault tolerance under {UNIX}.
\newblock {\em {ACM} Transactions on Computing Systems}, 7(1):1--24, 1989.

\bibitem{bottomley02clusters}
J.~Bottomley.
\newblock Clusters and high availability.
\newblock Presentation, January 2002.

\bibitem{candea04improving}
G.~Candea, J.~Cutler, and A.~Fox.
\newblock Improving availability with recursive micro-reboots: A soft-state
  system case study.
\newblock {\em Performance Evaluation Journal}, 56(1--3), March 2003.

\bibitem{candea01recursive}
G.~Candea and A.~Fox.
\newblock Recursive restartability: Turning the reboot sledgehammer into a
  scalpel.
\newblock In {\em 8th Workshop on Hot Topics in Operating Systems}, pages
  125--132, 2001.

\bibitem{ftpds04}
S.~Chakravorty and L.~V. Kale.
\newblock A fault tolerant protocol for massively parallel machines.
\newblock In {\em FTPDS Workshop for IPDPS 2004}. IEEE Press, 2004.

\bibitem{chandy85}
M.~Chandy and L.~Lamport.
\newblock Distributed snapshots: Determining global states of distributed
  systems.
\newblock {\em {ACM} Transactions on Computer Systems}, 31(1):63--75, 1985.

\bibitem{cristian91}
F.~Cristian and F.~Jahanian.
\newblock A timestamp-based checkpointing protocol for long-lived distributed
  computations.
\newblock In {\em Proceedings of the {IEEE} International Symposium on Reliable
  Distributed Systems}, pages 12--20, September 1991.

\bibitem{elnozahy94}
E.~N. Elnozahy and W.~Zwaenepoel.
\newblock {\em Manetho: fault tolerance in distributed systems using
  rollback-recovery and process replication}.
\newblock PhD thesis, Rice University, Department of Computer Science, 1994.

\bibitem{elnozahy-survey}
E.~N.~M. Elnozahy, L.~Alvisi, Y.-M. Wang, and D.~B. Johnson.
\newblock A survey of rollback-recovery protocols in message-passing systems.
\newblock {\em ACM Comput. Surv.}, 34(3):375--408, 2002.

\bibitem{fabre98metaobject}
J.-C. Fabre and T.~Perennou.
\newblock A metaobject architecture for fault-tolerant distributed systems: The
  {FRIENDS} approach.
\newblock {\em IEEE Transactions on Computers}, 47(1):78--95, 1998.

\bibitem{goldberg01design}
D.~Goldberg, M.~Li, W.~Tao, and Y.~Tamir.
\newblock The design and implementation of a fault-tolerant cluster manager.
\newblock Technical Report Computer Science Department Technical Report
  CSD-010040, University of California, Los Angeles, CA, October 2001.

\bibitem{hough00algorithm}
P.~D. Hough, M.~E. Goldsby, and E.~J. Walsh.
\newblock Algorithm-dependent fault tolerance for distributed computing.
\newblock Technical report, Sandia National Laborories, February 2000.

\bibitem{johnson87}
D.~B. Johnson and W.~Zwaenepoel.
\newblock Sender-based message logging.
\newblock In {\em The 17th Annual International Symposium on Fault-Tolerant
  Computing}, 1987.

\bibitem{kalbarczyk99chameleon}
Z.~T. Kalbarczyk, R.~K. Iyer, S.~Bagchi, and K.~Whisnant.
\newblock Chameleon: {A} software infrastructure for adaptive fault tolerance.
\newblock {\em IEEE Transactions on Parallel and Distributed Systems},
  10(6):560--588, 1999.

\bibitem{lamport78}
L.~Lamport.
\newblock Time, clocks, and the ordering of events in a distributed system.
\newblock {\em Communications of the {ACM}}, 21(7):565--588, 1978.

\bibitem{lamport82byzantine}
L.~Lamport, R.~Shostak, and M.~Pease.
\newblock The byzantine generals problem.
\newblock {\em ACM Transactions on Programming Languages and Systems},
  4(3):382--401, July 1982.

\bibitem{lamson79}
B.~Lamson and H.~Sturgis.
\newblock Crash recovery in a distributed data storage system.
\newblock Technical report, Xerox Palo Alto Research Center, 1979.

\bibitem{leangsuksun03building}
C.~Leangsuksun, T.~Liu, L.~Shen, and S.~L. Scott.
\newblock Building high availability and performance clusters with ha-oscar
  toolkits.
\newblock In {\em Proceedings of the High Availability and Performance
  Workshop}, October 2003.

\bibitem{avalon}
R.~A. Lerner.
\newblock Reliable servers: design and implementation in avalon/c++.
\newblock In {\em Proceedings of the first international symposium on Databases
  in parallel and distributed systems}, pages 13--21. IEEE Computer Society
  Press, 1988.

\bibitem{li01faulttolerant}
M.~Li, D.~Goldberg, W.~Tao, and Y.~Tamir.
\newblock Fault-tolerant cluster management for reliable high-performance
  computing.
\newblock In {\em International Conference on Parallel and Distributed
  Computing and Systems}, pages 480--485, 2001.

\bibitem{moser96totem}
L.~E. Moser, P.~M. Melliar-Smith, D.~A. Agarwal, R.~K. Budhia, and C.~A.
  Lingley-Papadopoulos.
\newblock {Totem}: {A} fault-tolerant multicast group communication system.
\newblock {\em Communications of the ACM}, 39(4):54--63, 1996.

\bibitem{arjuna}
G.~D. Parrington, S.~K. Shrivastava, S.~M. Wheater, and M.~C. Little.
\newblock The design and implementation of arjuna.
\newblock {\em Computing Systems}, 8(2):255--308, 1995.

\bibitem{pingali04application}
K.~Pingali.
\newblock Application-level fault tolerance for mpi programs.
\newblock Presentation, 8 2004.

\bibitem{powell93distributed}
D.~Powell.
\newblock Distributed fault-tolerance -- lessons learnt from delta-4.
\newblock In {\em Workshop on Fault-Tolerant Architectures}, page~18, Mont St.
  Michel, France, 1993. IRISA.

\bibitem{randell75}
B.~Randell.
\newblock System structure for software fault tolerance.
\newblock {\em {IEEE} Transactions on Software Engineering}, 1(2):220--232,
  1975.

\bibitem{reiter95rampart}
M.~K. Reiter.
\newblock The rampart toolkit for building high-integrity services.
\newblock In {\em Theory and Practice in Distributed Systems}, volume 938,
  pages 99--110. Springer-Verlag, Berlin Germany, 1995.

\bibitem{schlichting83}
R.~Schlichting and F.~Schneider.
\newblock Fail-stop processors: An approach to designing fault-tolerant
  computing systems.
\newblock {\em {ACM} Transactions on Computing Systems}, 1(3):222--238, 1983.

\bibitem{schulz04implementation}
M.~Schulz, G.~Bronevetsky, R.~Fernandes, D.~Marques, K.~Pingali, and
  P.~Stodghill.
\newblock Implementation and evaluation of a scalable application-level
  checkpoint-recovery scheme for mpi programs.
\newblock In {\em Proceedings of SC2004}, 2004.

\bibitem{shokri97roafts}
E.~Shokri, P.~Crane, J.~Dussault, K.~Kim, and C.~Subbaraman.
\newblock {ROAFTS}: A {CORBA}-based middleware for real-time object-oriented
  adaptive fault tolerance support.
\newblock In {\em {IEEE} Workshop on Middleware for Distributed Real-Time
  Systems and Services}, San Francisco, CA, December 1997.

\bibitem{stratus}
The stratus {ActiveService} architecture: Remote access to mission-critical
  support, 24/7.
\newblock Technical report, Stratus Technologies, 2004.

\bibitem{strom85}
R.~Strom and S.~Yemini.
\newblock Optimistic recovery in distributed systems.
\newblock {\em {ACM} Transactions on Computing Systems}, 3(3):204--226, 1985.

\bibitem{tong92}
Z.~Tong, R.~Kain, and W.~Tsai.
\newblock Rollback-recovery in distributed systems using loosely synchronized
  clocks.
\newblock {\em {IEEE} Transactions on Parallel and Distributed Systems},
  3(2):246--251, 1992.

\end{thebibliography}

\end{document}